![IntechOpen]

**Book Chapter Template**

# Excitons in two-dimensional materials

Xiaoyang Zheng, Xian Zhang*
Stevens Institute of Technology, Hoboken, New Jersey, United States
*E-mail address of the corresponding author: xzhang4@stevens.edu

## Abstract

Because of the reduced dielectric screening and enhanced Coulomb interactions, two-dimensional (2D) materials like phosphorene and transition metal dichalcogenides (TMDs) exhibit strong excitonic effects, resulting in fascinating many-particle phenomena covering both intralayer and interlayer excitons. Their intrinsic bandgaps and strong excitonic emissions allow the possibility to tune the inherent optical, electrical, and optoelectronic properties of 2D materials via a variety of external stimuli, making them potential candidates for novel optoelectronic applications. In this review, we summarize exciton physics and devices in 2D semiconductors and insulators, especially in phosphorene, TMDs, and their van der Waals heterostructures (vdWHs). In the first part, we discuss the remarkably versatile excitonic landscape, including bright and dark excitons, trions, biexcitons, and interlayer excitons. In the second part, we examine common control methods to tune excitonic effects via electrical, magnetic, optical, and mechanical means. In the next stage, we provide recent advances on the optoelectronic device applications, such as electroluminescent devices, photovoltaic solar cells, and photodetectors. We conclude with a brief discussion on their potential to exploit vdWHs towards unique exciton physics and devices.

## 1. Introduction

Since the first 'modern' 2D material, monolayer graphene, was mechanically exfoliated in 2004 [1], the family of 2D materials has been extensively flourishing, covering insulators, semiconductors, semimetals, metals, and superconductors (Figure 1). In addition to semimetal graphene, other actively researched 2D materials include wide-bandgap insulator hexagonal boron nitride (hBN) [2], direct bandgap semiconductor phosphorene [3], Xenes (e.g., Monolayers of silicon (silicene), germanium (germanene) and tin (stanene)) [4], and transition metal dichalcogenides (TMDs) with the chemical formula $MX_2$ (M: transition metal; X: chalcogen) [5]. Compared with bulk materials, 2D materials exhibit some unparallel characteristics: removal of van der Waals interactions, an increase in the ratio of surface area-to-volume, and confinement of electrons in a plane. The change in properties, caused by a reduction in the dimensionality of 2D materials, makes them becoming the promising candidates for next-generation electronics and optoelectronics [6].

Whereas these materials are marvelous per se, the more astounding discovery is that these 2D crystals can be combined freely to create layered compounds, paving a way for design of new functional materials and nano-devices [7,8]. Such designer materials are called van der Waals heterostructures (vdWHs) since the atomically thin layers are not attached through a chemical reaction but rather held together via a weak van der Waals interaction. By stacking together any number of atomically thin layers, the concept provides a huge potential to tailor the unique 2D electronic states with atomic scale precision, opening the door to broaden the versatility of 2D materials and devices. Such stacked vdWHs are quite distinctive from the traditional 3D semiconductor heterostructures, as each layer acts simultaneously as the bulk material and the interface, reducing the amount of charge displacement within each layer. These vdWHs have already gained an insight into the discovery of considerably engaging physical phenomena. For instance, by combining semiconducting monolayers with graphene, one can fabricate optically active heterostructures used for photovoltaic and light-emitting devices [9-11].

Because of the charge confinement and reduced dielectric screening, the optical properties of semiconducting 2D materials are dominated by excitonic effects [12-14]. When a material goes from bulk to 2D, there is less material to screen the electric field, giving rise to an increase in Coulomb interaction and more strongly-bound electron-hole pairs (excitons). In addition, since the excitons are confined in a plane that is thinner than their Bohr radius in most 2D semiconductors, quantum confinement enhances the exciton binding energy, altering the wavelength of light they absorb and emit. These two distinctively physical phenomena naturally make the excitons bound even at room temperature with a binding energy of hundreds of meV [15]. As a consequence, such materials' two-dimensionality makes the excitons easily tunable, with a variety of external stimuli or internal stacking layers, enabling them potential candidates for various applications in optics and optoelectronics.

In this chapter, we provide a topical summary toward recent frontier research progress related to excitons in atomically thin 2D materials and vdWHs. To begin with, we clarify the different types of excitons in 2D materials, including bright and dark excitons, trions, biexcitons, and interlayer excitons. Moreover, we analyze the electronic structures and excitonic effects for two typical 2D materials (i.e., TMDs and phosphorene), as well as the excited-state dynamics in vdWHs. Furthermore, we address how external stimuli, such applied electric fields, strain, magnetic fields, and light, modulate the excitonic behavior and emission in 2D materials. Afterward, we introduce several representative optoelectronic and photonic applications based on excitonic effects of 2D materials. Finally, we give our personal insights into the challenges and outlooks in this field.

## 2. Exciton physics in 2D semiconductors and insulators

When the dimension of crystals converts from 3D to 2D, the electronic Coulomb screening is dramatically reduced out of quantum confinement. As a consequence, dielectric constant $\epsilon$ can fall to $\epsilon=1$ from $\epsilon \gg 1$ in conventional bulk materials [16, 17]. Generally, the binding energies of the strongly bound excitons can reach up to 30% of the quasiparticle (QP) band gap because of the tremendous decrease in dielectric constant, rising to the magnitude of 0.1-1eV [15, 18]. The large binding energies, which lead to a strong absorption of excitons linking to light, can not only contribute to a substantial modification in the optical spectrum both below and above the QP band gap, but also ensure a long lifetime of excitons in room temperature. Since the large binding energies of excitons in 2D monolayer hBN was initially predicted theoretically in 2006 [19], the research relating to excitons of 2D materials boomed, ranging from monolayer 2D semiconductors and insulators to vdWHs.

**2.1 Excitons, trions, biexcitons, and interlayer excitons**

Excitons are hydrogen-like bound states of a negatively charged electron and a positively charged hole which are attracted to each other by the electrostatic Coulomb force [20]. It is an electrically neutral quasiparticle that exists mostly in semiconductors, as well as some insulators and liquids, derived from the photo-excitation. Excitons are the main mechanism for light emission and recombination because of their large oscillator strength and enhanced light-matter interaction [21]. When it comes to low-dimension crystals, the types of excitons experience a boom. Weak dielectric screening and strong geometrical confinement mutually contribute to an extremely strong Coulomb interaction, bringing in engaging many-particle phenomena: bright and dark excitons, trions, biexcitons, and interlayer excitons.

Excitons can be bright or dark subject to the spin orientation of the individual carriers: the electron and the hole, as shown in Figure 2b. If the electron and hole have opposite spins, the two particles can easily recombine through the emission of a photon. These electron-hole pairs are called bright excitons. Whereas if they have the same spins, the electron and hole cannot easily recombine via direct emission of a photon due to the lack of required spin momentum conservation. These electron-hole pairs are called dark excitons. This darkness makes dark excitons becoming promising qubits because dark excitons cannot emit light and are thus unable to relax to a lower energy level. As a consequence, dark excitons have relatively long radiative lifetimes, lasting for over a microsecond, a period that is a thousand times longer than bright excitons and long enough to function as a qubit. By harnessing the recombination time to create 'fast' or 'slow' light, the highly stable, non-radiative nature of dark excitons paves a way for optically controlled control information processing. For instance, according to inducing light emission from dark excitons in monolayer $WSe_2$, it is possible to selectively control spin and valley, making dark excitons possible to encode and transport information on a chip [22, 23].

Because of the significant Coulomb interactions in 2D materials, exciton can capture an additional charge to form charged exciton known as trion, a localized excitation consisting of three charged quasiparticles (Figure 2c). Compared to exciton, a neutral electron-hole pair, trion can be negative or positive depending on its charged state: a negative trion (negative e-e-h) is a complex of two electrons and one hole and a positive trion (negative e-e-h) is a complex of two holes and one electron. Trion states were predicted theoretically [24] and then observed experimentally in various 2D materials, by means of temperature-dependent photoluminescence (PL) [25] and nonlinear optical spectroscopy [26], and scanning tunneling spectroscopy [27]. Trions play a significant role in in manipulating electron spins and the valley degree of freedom for the reasons below. First, the trion binding energies are surprisingly large, reaching to about 15-45meV in monolayer TMDs [28-30] and 100meV in monolayer phosphorene on SiO2/Si substrate [31]. In addition, trions possess an extended population relaxation time up to tens of picoseconds [32, 33]. Finally, trions have an impact on both transport and optical properties and can be easily detected and tuned experimentally [34]. As a consequence, the electrical manipulation and detection of trion, as well as its enhanced stability, make it promising for trion-based optoelectronics.

Biexcitons, also known as exciton molecules, are created from two free excitons. Biexciton configurations can be distinguished from unbound or bound biexciton cases (Figure 2d). The bound biexciton is considered as a single particle since Coulomb interaction is dominant in this complex; while the unbound biexciton is regarded as two-exciton isolated from each other because of the predominance of the repulsive Coulomb interaction [35, 36]. Similar to trions stably existing in 2D materials, biexcitons can also exist in room temperature. Among 2D materials, biexcitons were firstly observed in monolayer TMDs [37, 38], followed by predicting their binding energies of biexcitons via computational simulation [39, 40].

In addition to above-mentioned intralayer excitons, interlayer excitons, where the involved electrons and holes are located in different layers, can also form in bilayer or few-layer 2D materials especially in vdWHs because of the strong Coulomb interaction (Figure 2e). After optically exciting a coherent intralayer exciton, the hole can tunnel to the other layer forming an incoherent exciton with the assistance of emission and absorption of phonons. Generally, these interlayer excitons occupy the energetically lower excitonic state than the excitons confined within one layer owing to an offset in the alignment of the monolayer band structures [41, 42]. Similar to excitons in one layer, interlayer excitons can also be either bright or dark depending on spin and momentum of the states involved [43, 44].

## 2.2 Excitons in atomically thin 2D materials

Among 2D semiconductors and insulators, TMDs and phosphorene have drawn tremendous attention owing to their intrinsic bandgaps and strong excitonic emissions, making them potential candidates for high-performance optoelectronic applications in the visible to near-infrared regime [46]. The electronic and optical properties of 2D materials rely on their electronic band structure, which demonstrates the movement of electrons in the material and results from the periodicity of its crystal structure. When the dimension of a material degrades from bulk to 2D, the periodicity will disappear in the direction perpendicular to the plane, changing the band structure dramatically. This means by changing the number of layers in the 2D material, one can tune the band structures (e.g., a $MoS_2$ will become emissive when reducing to monolayer), as well as tailor the binding energies of excitons (e.g., a monolayer 2D material will absorb/emit higher energy light than a bilayer).

All TMDs have a hexagonal structure, with each monolayer consisting of the metal layer sandwiched between two chalcogenide layers (X-M-X). The two most common crystal structures are the semiconducting 2H-phase with trigonal symmetry (e.g., $MoS_2$, $WS_2$, $MoSe_2$, $WSe_2$, as shown in Figure 3a) and the metallic 1T phase (e.g., $WTe_2$). For the semiconducting 2H-phase TMDs, they are well-known to possess an indirect band gap in bulk crystals; however, when mechanically exfoliated to a monolayer, these crystals experience a crossover from indirect to direct bandgap since the lack of interlayer interaction (Figure 3b). In addition, a decreasing layer numbers in TMDs attributes to larger absorption energy and strong photoluminescence (PL) emission in the visible spectrum, accompanying with enhanced excitonic effects, because of the reduced electronic Coulomb screening (Fig 4a) [47].

More importantly, TMDs are time-reversal symmetry but spatial inversion asymmetric. Since the strong spin–orbit coupling, the time-reversal symmetry dictates the spin splitting to have opposite spins at the K and K′ valleys of the Brillouin zones, making the excitons in TMDs are called valley excitons, which is different from the transition at the Γ valley in other 2D semiconductors such as phosphorene. As shown in Figure 5, the spin splitting is pretty strong in the valence band, in which spin splitting values are calculated theoretically up to 0.15 eV in 2H-MoS$_2$ monolayer and 0.46 eV in 2H-WSe$_2$ monolayer [48]. On the other hand, the broken inversion symmetry of TMD systems gives rise to a valley-dependent optical selection rule. This unique characteristic arouses the potential to control valley polarization and electronic valley. In this sense, a valley refers to the region in an electronic band structure where excitons are localized; valley polarization refers to the ratio of valley populations; and electronic valley refers a degree of freedom that is akin to charge and spin. As a consequence, optical transitions such as excitons in opposite valleys are able to be excited selectively using light with disparate chirality, paving the way to enable valleytronic devices based on photon polarizations [49-50].

As shown in figure 3c, phosphorene possesses a puckered orthorhombic lattice structure with P atoms distributed on two parallel planes and each P atom is covalently bonded to three adjacent atoms, resulting in strong in-plane anisotropy. Unlike TMDs that exhibit an indirect-to-direct bandgap transition when scaled down from bilayer to monolayer, phosphorene retains a direct band gap all the time, as shown in Figure 3d [51,52]. As the layer number decrease from 5 to 1, bandgap energy of phosphorene rises remarkably because of the weaker coupling of the conduction band and the valence band caused by reduced interactions in thinner layers, showing a layer-dependent direct bandgap energies (Figure 4c). In contrast to TMDs whose PL emission occurs in the visible spectrum, the light emission of phosphorene mainly covers the near-infrared spectral regime (Figure 4a and 4b). Moreover, its structural anisotropy also strongly affects the excitonic effects and in phosphorene. The results from first-principles simulations demonstrate that excitonic effects can only be observed when the incident light is polarized along the armchair direction of the crystal [53].

To have an impact on excitonic effects and relevant applications, the binding energy of these quasiparticles must be clarified. As schematically illustrated in Figure 2a, the exciton binding energy is the energy difference between the electronic bandgap ($E_g$) and optical bandgap ($E_{opt}$). When higher-order excitonic quasiparticles form, more energy, i.e., the binding energy of trion or biexciton, is needed. Thus, the binding energies of exciton, trion and biexciton can be expressed as $E_b^E = E_g - E_E, E_b^T = E_g - E_T$, and $E_b^B = E_g - E_B$, respectively, where $E_E$, $E_T$, and $E_B$ are emission energies of exciton, trion, and biexciton. For the most 2D conductors and insulators, a robust linear scaling law exists between the quasiparticle bandgap ($E_g$) and the exciton binding energy ($E_b^E$), namely, $E_b^E \approx E_g/4$, regardless of their lattice configuration, bonding characteristic, and the topological property (Figure 6) [54]. It is worth emphasizing that the results from simulations and experiments cover almost all kinds of popular 2D monolayer semiconductors and insulators, including topological crystalline insulator (TCI) and topological insulator (TI) [55-57], TMDs [54, 58-61], nitrides (MXenes) [62], phosphorene [54, 63], IV/III–V compounds [54], and graphene derivatives [54]. Such an agreement between simulation and experiment results indicates that the linear scaling law can be used effectively to predict the exciton binding energy for all the 2D monolayer semiconductors and insulators. On the other hand, although comparatively lower than exciton binding energies, the binding energies of trion and biexciton in 2D materials is significantly larger than that in quasi-2D quantum wells (1-5 meV) [64]. For example, the binding energies of trion and biexciton in TMDs reach up to 45 meV and 60 meV, respectively [36, 38, 65].

**2.3 Excitons in vdWHs**

Composed of stacks of atomically thin 2D materials, the properties of vdWHs are determined not only by the constituent monolayers but also by the layer interactions. In particular, the excited-state dynamics is unique, such as the formation of interlayer excitons [41], ultrafast charge transfer between the layers [66, 67], the existence of long-lived spin and valley polarization in resident carriers [68,69], and moiré-trapped valley excitons in moire superlattices in vdWHs [70-73]. In terms of 2D vdWHs, the semiconducting vdWHs composed of stacked TMDC layers are the most widely studied due to their prominent exciton states and accessibility to the valley degree of freedom. More interestingly, the introduction of moiré superlattices (Figure 7a), a periodic pattern formed by stacking two monolayer 2D materials with lattice mismatch or rotational misalignment, enables to modulate the electronic band structure and the optical properties of vdWHs [74].

After demonstrating the appearance of interlayer excitons in PL spectra, the research on exciton dynamics in vdWHs flourishes. The discovery of intralayer excitons in 2D materials can be traced back to 2015, when long-lived interlayer excitons were demonstrated in monolayer $MoSe_2/Wse_2$ heterostructures, where a pronounced additional resonance was observed at an energy below the intralayer excitons [75]. Compared with the intralayer excitons in the weak excitation regime, the PL intensity of this low-energy peak is rather prominent, which attributes to the presence of interlayer excitons as their spectral position are highly occupied. Furthermore, measuring the binding energy of interlayer excitons directly is also demonstrated in $WSe_2/WS_2$ heterobilayers, where a novel 1s–2p resonance are measured by phase-locked mid-infrared pulses [76]. For other excited-state dynamics, such as ultrafast kinetics, long lifetimes, and moiré excitons, some research indicate they have something to do with interlayer excitons [66-73].

Empirically, charge transfer between layers of vertically stacking vdWHs is supposed to be much slow. However, transient absorption measurements, which are implemented by resonantly injecting excitons using ultrafast laser pulse, show a sub-picosecond charge separation in vdWHs: the holes injected in $MoS_2$ takes 200 fs transferring to $MoSe_2$ and even only 50 fs transferring to $WS_2$, as shown in Figure 7b [68,69]. It is noteworthy that this process is reversible, i.e., holes transfer to $MoSe_2$ on the same ultrafast time scale when excitons are selectively injected in $MoS_2$ using excitation resonant with the higher-energy exciton feature in $MoS_2$. In addition, another interesting phenomenon is that when mismatching the bilayer vdWHs with different twist angle, the charge transfer signal keeps a constant period within 40 fs, while the recombination lifetime of these indirect excitons varies with the twist angle without any clear trend [77].

In contrast to the ultrafast charge transfer dynamics in vdWHs (<1ps), spin and valley relaxation dynamics take place on considerably longer timescale [68,69]. For the two distinctive relaxation processes in vdWHs (i.e., the population decay of optically excited excitons, and the exciton spin–valley lifetime which determines the information storage time in the spin– valley degree of freedom), they both are significantly longer than the monolayer case. For instance, by tuning the carrier concentration, holes' spin–valley lifetime and population lifetime possess a doping-dependent pattern in a $WSe_2/WS_2$ heterostructure [69]: in charge-neutral and electron-doped heterostructures (i.e., neutral and positive carrier concentrations), the spin–valley lifetime is closed to the population lifetime; nevertheless, in hole-doping heterostructures (i.e., negative carrier concentration), the spin–valley lifetime becomes orders of magnitude longer than the population lifetime (Figure 7c). The remarkable dynamics of doping-dependent lifetime attributes to the distinctive interlayer electron-hole recombination process in the heterostructure, as shown in Figure 7c. In electron-doped or charge-neutral heterostructures, all holes in $WSe_2$ are pump-generated excess holes; hence, when the hole population decays to zero out of interlayer electron-hole recombination, no holes can remain, let alone valley-polarized holes. The valley lifetime is thus limited by the lifetime of the total hole excess. On the contrary, in hole-doped case, the original hole density is much higher than the photo-generated density, give an equal probability for the recombination of excess electrons in $WS_2$ with holes from both valleys of $WSe_2$.

Since 2019, important breakthroughs about excitons in vdWHs has been obtained, especially three independent research simultaneously reporting the observation of moiré excitons in TMDs vdWHs, which lays a firm foundation to the engineering artificial excitonic crystals using vdWHs for nanophotonics and quantum information applications [70-72]. For example, in $MoSe_2/WSe_2$ heterobilayers with a small twist angle of ~1°, there are three points at which the local atomic registration preserves the threefold rotational symmetry $\hat{C}_3$ in the moiré supercell. The local energy extrema in the three high-symmetry points not only localizes the excitons but also provides an array of identical quantum-dot potentials (Figure 7d) [70]. The research on moiré excitons in TMDs vdWHs has been promoted after experimentally confirming the hybridization of excitonic bands that can result in a resonant enhancement of moiré superlattice effects.

## 3. Tuning methods of excitons

To have an impact on industrial applications especially photovoltaics, the binding energies of excitons in 2D semiconductors and insulators must be delicately designed and tuned. More importantly, these common control measures, from electrical to optical methods, function more potently in 2D materials than in 3D materials.

### 3.1 Electrical tuning

Since the electric field can hardly modulate the dielectric constant in monolayer 2D materials [78], early electrical tuning for excitonic behavior is mostly based on carrier density-dependent many-body Coulomb interactions, namely charged excitons or trions [79,80]. By increasing electron doping density using different gate voltage (-100 to +80 V) in monolayer $MoS_2$ field-effect transistors, Mak et al. firstly reported the observation of tightly bound negative trions by means of absorption and photoluminescence spectroscopy [79]. These negative trions hold a large trion binding energy up to ~20 meV, and can be optically created with valley and spin polarized holes. At the same time, Ross et al. also observed positive and negative trions along with neutral excitons in monolayer $MoSe_2$ field-effect transistors via photoluminescence [80]. The exciton charging effects shown a reversible electrostatic tunability, as shown in Figure 8a-8c. More interestingly, the positive and negative trions exhibited a nearly identical binding energy (~30 meV), implying the same effective mass for electrons and holes. Another work demonstrated continuous tuning of the exciton binding energy in monolayer $WS_2$ field-effect transistors, finding the ground and excited excitonic states as a function of gate voltage [81].

The above-mentioned works are related to monolayer 2D materials, while when it comes to heterostructures, the electrical tuning functions more efficiently [82,83]. Employing a van der Waals heterostructure consisting of $hBN/MoSe_2/hBN$ (Figure 8d and 8e), Wang el al. obtained homogeneous 2D electron gases by controlling disorder in TMDs, which allows for excellent electrical control of both charge and excitonic degrees of freedom [82]. Measuring the optoelectronic transport in the gate-defined heterostructure, they demonstrated gate-defined and tunable confinement of charged exciton, i.e., confinement happens when local gate voltages $\Delta V_g$ is zero or negative while being absent when $\Delta V_g$ local gate voltages are positive (Figure 8f). To further demonstrate controlled localization of charged excitons, they excited the device with a laser source at λ = 660 nm, observing both the exciton and trion recombination in PL spectra (Figure 8g). The ratio between trion and exciton recombination emission declines as $\Delta V_g$ becomes more negative, because of local depletion of trions as the device transits from the accumulation regime ($\Delta V_g>0$) to confinement ($\Delta V_g=0$) and depletion regimes ($\Delta V_g<0$), respectively.

### 3.2 Magnetic tuning

TMDs have drawn more attention with respect to magnetic tuning than other 2D materials, since they preserve time-reversal symmetry with excitons formed at K and K′ points at the boundary of the Brillouin zone, which restricts valley polarization. However, when imposing magnetic fields, time-reversal symmetry can be broken, which splits the degeneracy between the nominally time-reversed pairs of exciton optical transitions at K and K′ valley: this is the valley Zeeman effect, as shown in Figure 9a and 9b [84–88]. Based on the Zeeman effect, magnetic manipulation is effectively used on valley pseudospin [84], valley splitting and polarization [85], and valley angular momentums [86]. For high-order excitonic quasiparticles, valley Zeeman effect also exhibit significant effects on trions [88] and biexcitons [89] under applied magnetic fields.

In addition, magnetic fields, which changes the surrounding dielectric environment, can also have an impact on the size and binding energy of excitons. By encapsulating the flakes with different materials on a monolayer $WSe_2$, Stier et al. changed the average dielectric constant, $k=(\varepsilon_t+\varepsilon_b)/2$, ranging from 1.55 to 3.0 (Figure 9c) [90]. The average energy of the field-split exciton transitions was measured in pulsed magnetic fields to 65 T, exhibiting an increasing trend with field which reveals the diamagnetic shift can infer both exciton binding energy and radius. They demonstrated increased environmental screening will enlarge exciton size but reduce exciton binding energy in 2D semiconductors, which shows a quantitatively agreement with theoretical models (Figure 9d).

### 3.3 Optical tuning

To control excitonic effects by breaking time-reversal symmetry in TMDs, imposing an intense circularly polarized light can also achieve the aim based on optical Stark effect, a phenomenon that photon-dressed states (Floquet states) can hybridize with the equilibrium states resulting in energy repulsion between the two states [91,92], as shown in Figure 10a and 10b. The interaction between Floquet and equilibrium states can not only bring in a wider energy level separation, but also enhance the magnitude of the energy repulsion if they are energetically close. Based on the optical Stark effect triggered off by circularly polarized light, two independent works demonstrated that the exciton level in K and K' valleys can be selectively tuned by as much as 18meV in $WS_2$ monolayer and 10meV in $WSe_2$ monolayer, respectively.

Besides, optical control and manipulation have been shown effective towards valley polaritons, a half-light half-matter quasiparticles arising from hybridization of an exciton mode and a cavity mode. Owing to the large exciton binding energy and oscillator strength in TMDs, spin–valley coupling can persist at room temperature when excitons are coherently coupled to cavity photons, leading to a stable exciton-polariton formation [93-96]. Exciton polaritons are interacting bosons with very light mass, and can be independently combined in the intracavity and extracavity field. A schematic of the valley-polariton phenomena is shown in Figure 10c, where the microcavity structure consists of silver mirrors with a silicon dioxide cavity layer embedded with the $WS_2$ monolayer. The valley-polarized exciton–polaritons are optical pumped using two pumps to excite the exciton reservoir and the lower polariton branch, showing an angle-dependent helicity because of the excitonic component of the polariton states [95]. In addition, another work based on similar method demonstrates that exciton-polaritons possess a temperature-dependent emission polarization, exhibiting stronger valley polarization at room temperature compared with bare excitons [96], as shown in Figure 10d.

### 3.4 Mechanical tuning

2D materials possess excellent mechanical flexibility, making them stable under high compressive, tensile, and bending strain [97]. Applying mechanical strain on 2D materials, their bandgaps will reduce, increase, or transit from direct to indirect, thus resulting in a strain-dependent exciton binding energy [98-103]. Based on density functional theory, Su et al. investigated the natural physical properties of TMD monolayers and hBN- TMD heterostructures, finding that they have distinctive bandgap and exciton binding energy under compressive strain (Figure 11a) [98]. $MS_2$ monolayers exhibit direct-to-indirect transition, while hBN-TMD heterostructures keep direct band-gap characters because of the strong charge transfer between hBN and TMD monolayers. With increasing compressive strain, the exciton binding energies of TMD monolayers gradually reduce, but the binding energies of hBN- TMD heterostructures experience a dramatically growth before decreasing (Figure 11b).

Another mechanical tuning method is by implying heterogeneous strain on 2D materials, which would result in a spatially varying bandgap with tunable exciton binding energy distribution, namely funnel effect [104-106]. In a TMDs monolayer, excitons will move towards high tensile strain region, resulting in a funnel-like band energy profile. In contrast, excitons in phosphorene are pushed away from high tensile strain region, exhibits inverse funnel effect of excitons, which is moreover highly anisotropic with more excitons flowing along the armchair direction (Figure 11c) [107]. Funnel effect is a rare method for control exciton movement, paving a way for creating a continuously varying bandgap profile in an initially homogeneous, atomically thin 2D materials.

## 4. Optoelectronic devices

2D materials possess strong light-matter coupling and direct band gaps from visible to infrared spectral regimes with strong excitonic resonances and large optical oscillation strength. Recent observation of valley polarization, exciton–polaritons, optically pumped lasing, exciton–polaritons, and single-photon emission highlights the potential for 2D materials for applications in novel optoelectronic devices. Combined with external stimuli, like electrical and magnetic fields, optical pumps, and strain, exciton effects in 2D materials show a highly tunability and flexibility in electroluminescent devices, photovoltaic solar cells, and photodetectors.

### 4.1 Electroluminescent devices

Excitonic electroluminescence (EL) emission in 2D materials is key to fully exploiting the EL devices [49, 108]. Based on different carrier injection and transport mechanisms, light-emitting devices have distinctive structures, depending on their mechanism of exciton generation: bipolar carrier injection in p-n heterojunction [109,110], quantum well heterostructures [111], unipolar injection [112], impact excitation [113], thermal excitation [114], and interlayer excitons [115] (Figure 12).

As exciton emission induced by bipolar carrier injection, p-n heterojunction is the simplest device to achieving EL. Depending on the contacting way the two monolayers connect, it can be vertical or lateral. Typical p-n junction is $MS_2/MSe_2$ heterostructure, since their counterparts lack the caliber to function effectively [116,117]. For instance, $MoS_2$ and $WS_2$, in which sulfur vacancies act as electron donors, are often naturally n-type; while $WSe_2$ and $MoSe_2$ are typically ambipolar but often unconsciously p-doped by adsorbed moisture [118,119].

Quantum well (QW) heterostructures consist of semiconductor layer sandwiched between insulator layers and metal electrodes. EL in QW heterostructures can be observed by bipolar recombination of injected electrons and holes in the semiconductor layer when applied a bias in the metal electrodes. Since the long lifetime of carriers and enhanced exciton formation in the semiconductor layer (typical one is TMDs), the emission efficiency of multiple QW devices is much higher than that of single QW devices, and can be improved by preparing alternating layers of TMDs [111,120].

Unipolar injection happens in a metal–insulator–semiconductor (MIS) or a semiconductor–insulator–semiconductor (SIS) heterostructures when a positive bias applied to the metal and semiconductor layers. The common insulator layer is hBN since its ability to transport holes but block electrons. If the bias is increased above a threshold, EL will be observed at extremely low current densities below 1 nA μm$^{-2}$, attributed to the unipolar tunneling across the hBN layer, which transfers holes from one metal/semiconductor layer (e.g., graphene and TMDs) to another electron-rich semiconductor layer (e.g., TMDs) [121].

The remaining three emission mechanism is relatively simple. For impact excitation devices, excitons are generated by impact excitation of excitons in the high field regime rather than bipolar recombination. For thermal emission devices, a semiconductor monolayer or few layers are partly suspended on a substrate, and thermal excitation and emission are evoked by locally heating the high current density regime. For bilayer emission devices, emission occurs due to the recombination of electrons and holes residing in the adjacent layers.

**4.2 Photovoltaic solar cells**

2D materials possess large exciton binding energy with the bandgap ranging from visible to near-infrared part of the spectrum, making them attractive as candidates for photovoltaic solar cells [122,123]. Light absorption in the active layers of a photovoltaic cell significantly determines device efficiency. To improve light absorption of 2D semiconductor photovoltaics in the ultrathin limit, light trapping designs are need, such as use of plasmonic metal particles, shells, or resonators to amplify photocurrent and photoluminescence. For large area photovoltaic applications, a common strategy is thin film interference, in which a highly reflective metal (e.g., Au or Ag) is used as a part of an "open cavity" to enhance absorption due to multipass light interactions within the semiconductor (Figure 13a) [124]. If the semiconductor layer is a monolayer absorber, an atomically thin absorber with $\lambda/4$ in thickness can be sandwiched between conductor layer and reflector layer, enabling destructive interference at the interface and thus resulting in significant absorption enhancement (Figure 13b) [125]. Another strategy to enhance light trapping is by the use of nanostructured resonators, which are coupled to or etched in thin film absorbers (Figure 13c and 13d) [126,127].

Compared with free-standing monolayer with merely 10% absorption [128], the above-mentioned strategy exhibits outstanding strength for TMDC devices. For example, TMD-reflector coupled photovoltaics can have high broadband absorption of 90% and quantum efficiency of 70% [129,130]. Accompanied with reflector, resonator, or antennas, 2D semiconductor photovoltaics trapping nearly 100% of the incident light may be achieved for nanoscale thick active layers. However, improving light absorption in sub-nanoscale thick monolayers faces more challenging, because not only the low absorption of monolayer (~10%) but also the limited technique to fabricate nanoscopic resonators or antennas [131].

**4.3 Photodetectors**

Photodetection is a process converting light signals to electric signals, consisting of three physical mechanisms: light harvesting, exciton separation, and charge carrier transport to respective electrodes. According to the operation modes, photodetectors can be divided into two categories: photoconduction (i.e., photoconductor) and photocurrent (i.e., photodiode) [132,133]. The former one refers to the overall conductivity change out of photoexcited carriers, and the latter one involves a junction which convert photoexcited carriers into current. Generally, photoconduction-based devices possess higher quantum efficiency than photocurrent-based devices, since transporting carriers can circulate many times before recombination in photoconductors. However, the response in photocurrent-based devices is faster than that in photoconduction-based devices, because of the short carrier lifetime that transporting carriers (electrons and holes) are both involved in the photocurrent generation and recombine with their counterpart after reaching to their own electrodes.

Two common 2D materials used for photodetectors are graphene [134-136] and TMDs (Figure 14) [50, 137, 138]. Based on photothermal with weak photovoltaic effect, graphene photodetectors usually show higher dark currents and smaller responsivity, but much wider operational bandwidths. In contrast, TMDs photodetectors operating on photovoltaic effect, exhibiting lower dark currents and higher responsivity.

For graphene-based photoconductors, typical devices are hybrid, adding a light absorption material, such as quantum dots [139], perovskites [140], silicon [141], carbon nanotubes [142], and TMDs [143], as active layer to improve the responsivity. For graphene-based photodiodes, this earliest reported one is metal–graphene–metal photodiodes, in which photocurrent was generated by local illumination of the metal/graphene interfaces of a back-gated graphene field-effect transistor. The resulting current can be attributed to either photovoltaic effect [144] or photo-thermoelectric effect [145]. To additionally improve the performance, common structures are graphene–semiconductor heterojunction photodiodes, in which planar junctions of graphene and group-IV elements or other compound semiconductors act as Schottky diodes [146, 147].

For TMD photodetectors, devices can have in-plane or out-of-plane structures, based on the semiconductor layers stacking laterally or vertically. In-plane devices take advantage of better control of the material's properties via electrostatic gating [148]. But out-of-plane devices can bear a much higher bias field (up to ~1 V nm$^{-1}$), enabling a reduced excitonic binding energy in multilayer structures for more efficient exciton dissociations [149]. TMDs-based photoconductors are usually enhanced by illuminating the semiconductor–metal contacts [150] and in short-channel devices [151], and their conductance can be changed by doping and trapping of photogenerated carriers by impurity states [152, 153]. On the other hand, TMDs-based photodiodes exhibit higher tunability based on the photocurrent mode, consisting of an in-plane or out-of-plane junction where a built-in electric field is created [154-156]. In this situation, electrostatic gates can further tune the device doping levels, owing to the very small interlayer separation (<1 nm) which produces extremely high built-in electric fields (~1 V nm$^{-1}$).

## 5. Summary and perspective

In summary, 2D materials exhibit excitonic effects due to spatial confinement and reduced screening at the 2D limit, resulting in fascinating many-particle phenomena, such as excitons, trions, biexcitons, and interlayer excitons. Enhanced binding energies owing to the strong Coulomb interaction makes these quasiparticles easy to characterize and control. In addition, the sensitivity of these quasiparticles to a variety of external stimuli allows the possibility of modulating the inherent optical, electrical, and optoelectronic properties of 2D materials, making them potential candidates for novel optoelectronic applications.

In addition to well-studied 2D materials, such as graphene, phosphorene, and TMDs, the family of 2D crystals is continuously growing, making excitonic effects versatile in different 2D systems. In particular, assembling vdWHs, which now can be mechanically assembled or grown by ample methods, can open up a new route for exploring unique exciton physics and applications. For example, an in-plane moiré superlattice, formed by vertically stacking two monolayer semiconductors mismatching or rotationally misaligning, can modulate the electronic band structure and thus lead to electronic phenomena, such as fractal quantum Hall effect, unconventional superconductivity, and tunable Mott insulators.

## Acknowledgments

This work is supported by the start-up funds at Stevens Institute of Technology.

## Conflict of Interest

The authors declare no conflict of interest.

## References

[1] Novoselov KS, Geim AK, Morozov SV, Jiang D, Zhang Y, Dubonos SV, Grigorieva IV, Firsov AA. Electric field effect in atomically thin carbon films. science. 2004 Oct 22;306(5696):666-9.

[2] Jin C, Lin F, Suenaga K, Iijima S. Fabrication of a freestanding boron nitride single layer and its defect assignments. Physical review letters. 2009 May 14;102(19):195505.

[3] Liu H, Neal AT, Zhu Z, Luo Z, Xu X, Tománek D, Ye PD. Phosphorene: an unexplored 2D semiconductor with a high hole mobility. ACS nano. 2014 Mar 21;8(4):4033-41.

[4] Molle A, Goldberger J, Houssa M, Xu Y, Zhang SC, Akinwande D. Buckled two-dimensional Xene sheets. Nature materials. 2017 Feb;16(2):163.

[5] Manzeli S, Ovchinnikov D, Pasquier D, Yazyev OV, Kis A. 2D transition metal dichalcogenides. Nature Reviews Materials. 2017 Aug;2(8):17033.

[6] Mas-Balleste R, Gomez-Navarro C, Gomez-Herrero J, Zamora F. 2D materials: to graphene and beyond. Nanoscale. 2011;3(1):20-30.

[7] Geim AK, Grigorieva IV. Van der Waals heterostructures. Nature. 2013 Jul;499(7459):419.

[8] Novoselov KS, Mishchenko A, Carvalho A, Neto AC. 2D materials and van der Waals heterostructures. Science. 2016 Jul 29;353(6298):aac9439.

[9] Furchi MM, Zechmeister AA, Hoeller F, Wachter S, Pospischil A, Mueller T. Photovoltaics in Van der Waals heterostructures. IEEE Journal of Selected Topics in Quantum Electronics. 2016 Jun 20;23(1):106-16.

[10] Furchi MM, Höller F, Dobusch L, Polyushkin DK, Schuler S, Mueller T. Device physics of van der Waals heterojunction solar cells. npj 2D Materials and Applications. 2018 Feb 19;2(1):3.

[11] Liu CH, Clark G, Fryett T, Wu S, Zheng J, Hatami F, Xu X, Majumdar A. Nanocavity integrated van der Waals heterostructure light-emitting tunneling diode. Nano letters. 2016 Dec 12;17(1):200-5.

[12] Pei J, Yang J, Yildirim T, Zhang H, Lu Y. Many‐Body Complexes in 2D Semiconductors. Advanced Materials. 2019 Jan;31(2):1706945.

[13] Xiao J, Zhao M, Wang Y, Zhang X. Excitons in atomically thin 2D semiconductors and their applications. Nanophotonics. 2017 Nov 1;6(6):1309-28.

[14] Mueller T, Malic E. Exciton physics and device application of two-dimensional transition metal dichalcogenide semiconductors. npj 2D Materials and Applications. 2018 Sep 10;2(1):29.

[15] Jiang Z, Liu Z, Li Y, Duan W. Scaling universality between band gap and exciton binding energy of two-dimensional semiconductors. Physical review letters. 2017 Jun 27;118(26):266401.

[16] Laturia A, Van de Put ML, Vandenberghe WG. Dielectric properties of hexagonal boron nitride and transition metal dichalcogenides: from monolayer to bulk. npj 2D Materials and Applications. 2018 Mar 8;2(1):6.

[17] Hwang EH, Sarma SD. Dielectric function, screening, and plasmons in two-dimensional graphene. Physical Review B. 2007 May 11;75(20):205418.

[18] Kidd DW, Zhang DK, Varga K. Binding energies and structures of two-dimensional excitonic complexes in transition metal dichalcogenides. Physical Review B. 2016 Mar 18;93(12):125423.

[19] Wirtz L, Marini A, Rubio A. Excitons in boron nitride nanotubes: dimensionality effects. Physical review letters. 2006 Mar 30;96(12):126104.

[20] Elliott RJ. Intensity of optical absorption by excitons. Physical Review. 1957 Dec 15;108(6):1384.

[21] Citrin DS. Radiative lifetimes of excitons in quantum wells: Localization and phase-coherence effects. Physical Review B. 1993 Feb 15;47(7):3832.

[22] Zhang XX, Cao T, Lu Z, Lin YC, Zhang F, Wang Y, Li Z, Hone JC, Robinson JA, Smirnov D, Louie SG. Magnetic brightening and control of dark excitons in monolayer WSe 2. Nature nanotechnology. 2017 Sep;12(9):883.

[23] Zhou Y, Scuri G, Wild DS, High AA, Dibos A, Jauregui LA, Shu C, De Greve K, Pistunova K, Joe AY, Taniguchi T. Probing dark excitons in atomically thin semiconductors via near-field coupling to surface plasmon polaritons. Nature nanotechnology. 2017 Sep;12(9):856.

[24] Ganchev B, Drummond N, Aleiner I, Fal'ko V. Three-particle complexes in two-dimensional semiconductors. Physical review letters. 2015 Mar 11;114(10):107401.

[25] Sun H, Wang J, Wang F, Xu L, Jiang K, Shang L, Hu Z, Chu J. Enhanced exciton emission behavior and tunable band gap of ternary W ($S_xSe_{1-x}$) 2 monolayer: temperature dependent optical evidence and first-principles calculations. Nanoscale. 2018;10(24):11553-63.

[26] Ye J, Yan T, Niu B, Li Y, Zhang X. Nonlinear dynamics of trions under strong optical excitation in monolayer MoSe 2. Scientific reports. 2018 Feb 5;8(1):2389.

[27] Demeridou I, Paradisanos I, Liu Y, Pliatsikas N, Patsalas P, Germanis S, Pelekanos NT, Goddard III WA, Kioseoglou G, Stratakis E. Spatially selective reversible charge carrier density tuning in $WS_2$ monolayers via photochlorination. 2D Materials. 2018 Oct 17;6(1):015003.

[28] Zhang DK, Kidd DW, Varga K. Excited biexcitons in transition metal dichalcogenides. Nano letters. 2015 Oct 2;15(10):7002-5.

[29] Berkelbach TC, Hybertsen MS, Reichman DR. Theory of neutral and charged excitons in monolayer transition metal dichalcogenides. Physical Review B. 2013 Jul 25;88(4):045318.

[30] Cadiz F, Tricard S, Gay M, Lagarde D, Wang G, Robert C, Renucci P, Urbaszek B, Marie X. Well separated trion and neutral excitons on superacid treated $MoS_2$ monolayers. Applied Physics Letters. 2016 Jun 20;108(25):251106.

[31] Yang J, Xu R, Pei J, Myint YW, Wang F, Wang Z, Zhang S, Yu Z, Lu Y. Optical tuning of exciton and trion emissions in monolayer phosphorene. Light: Science & Applications. 2015 Jul;4(7):e312.

[32] Gao F, Gong Y, Titze M, Almeida R, Ajayan PM, Li H. Valley trion dynamics in monolayer $MoSe_2$. Physical Review B. 2016 Dec 12;94(24):245413.

[33] Wang G, Bouet L, Lagarde D, Vidal M, Balocchi A, Amand T, Marie X, Urbaszek B. Valley dynamics probed through charged and neutral exciton emission in monolayer $WSe_2$. Physical Review B. 2014 Aug 18;90(7):075413.

[34] Mak KF, McGill KL, Park J, McEuen PL. The valley Hall effect in $MoS_2$ transistors. Science. 2014 Jun 27;344(6191):1489-92.

[35] Sahin M, Koç F. A model for the recombination and radiative lifetime of trions and biexcitons in spherically shaped semiconductor nanocrystals. Applied Physics Letters. 2013 May 6;102(18):183103.

[36] Steinhoff A, Florian M, Singh A, Tran K, Kolarczik M, Helmrich S, Achtstein AW, Woggon U, Owschimikow N, Jahnke F, Li X. Biexciton fine structure in monolayer transition metal dichalcogenides. Nature Physics. 2018 Dec;14(12):1199.

[37] You Y, Zhang XX, Berkelbach TC, Hybertsen MS, Reichman DR, Heinz TF. Observation of biexcitons in monolayer $WSe_2$. Nature Physics. 2015 Jun;11(6):477.

[38] Shang J, Shen X, Cong C, Peimyoo N, Cao B, Eginligil M, Yu T. Observation of excitonic fine structure in a 2D transition-metal dichalcogenide semiconductor. ACS nano. 2015 Jan 12;9(1):647-55.

[39] Mayers MZ, Berkelbach TC, Hybertsen MS, Reichman DR. Binding energies and spatial structures of small carrier complexes in monolayer transition-metal dichalcogenides via diffusion Monte Carlo. Physical Review B. 2015 Oct 9;92(16):161404.

[40] Szyniszewski M, Mostaani E, Drummond ND, Fal'Ko VI. Binding energies of trions and biexcitons in two-dimensional semiconductors from diffusion quantum Monte Carlo calculations. Physical Review B. 2017 Feb 7;95(8):081301.

[41] Ovesen S, Brem S, Linderälv C, Kuisma M, Korn T, Erhart P, Selig M, Malic E. Interlayer exciton dynamics in van der Waals heterostructures. Communications Physics. 2019 Feb 28;2(1):23.

[42] Ross JS, Rivera P, Schaibley J, Lee-Wong E, Yu H, Taniguchi T, Watanabe K, Yan J, Mandrus D, Cobden D, Yao W. Interlayer exciton optoelectronics in a 2D heterostructure p–n junction. Nano letters. 2017 Jan 4;17(2):638-43.

[43] Chen Y, Quek SY. Tunable bright interlayer excitons in few-layer black phosphorus based van der Waals heterostructures. 2D Materials. 2018 Sep 25;5(4):045031.

[44] Latini S, Winther KT, Olsen T, Thygesen KS. Interlayer excitons and band alignment in $MoS_2$/hBN/$WSe_2$ van der Waals heterostructures. Nano letters. 2017 Jan 4;17(2):938-45.

[45] https://www.ossila.com/pages/graphene-2d-materials [Accessed: 2019-07-14]

[46] Yu S, Wu X, Wang Y, Guo X, Tong L. 2D materials for optical modulation: challenges and opportunities. Advanced Materials. 2017 Apr;29(14):1606128.

[47] Zhao W, Ghorannevis Z, Chu L, Toh M, Kloc C, Tan PH, Eda G. Evolution of electronic structure in atomically thin sheets of $WS_2$ and $WSe_2$. ACS nano. 2012 Dec 28;7(1):791-7.

[48] Zhu ZY, Cheng YC, Schwingenschlögl U. Giant spin-orbit-induced spin splitting in two-dimensional transition-metal dichalcogenide semiconductors. Physical Review B. 2011 Oct 14;84(15):153402.

[49] Mak KF, Shan J. Photonics and optoelectronics of 2D semiconductor transition metal dichalcogenides. Nature Photonics. 2016 Apr;10(4):216.

[50] Wang J, Verzhbitskiy I, Eda G. Electroluminescent devices based on 2D semiconducting transition metal dichalcogenides. Advanced Materials. 2018 Nov;30(47):1802687.

[51] Zhang S, Yang J, Xu R, Wang F, Li W, Ghufran M, Zhang YW, Yu Z, Zhang G, Qin Q, Lu Y. Extraordinary photoluminescence and strong temperature/angle-dependent Raman responses in few-layer phosphorene. ACS nano. 2014 Sep 8;8(9):9590-6.

[52] Qiao J, Kong X, Hu ZX, Yang F, Ji W. High-mobility transport anisotropy and linear dichroism in few-layer black phosphorus. Nature communications. 2014 Jul 21;5:4475.

[53] Tran V, Soklaski R, Liang Y, Yang L. Layer-controlled band gap and anisotropic excitons in few-layer black phosphorus. Physical Review B. 2014 Jun 26;89(23):235319.

[54] Jiang Z, Liu Z, Li Y, Duan W. Scaling universality between band gap and exciton binding energy of two-dimensional semiconductors. Physical review letters. 2017 Jun 27;118(26):266401.

[55] Xu Y, Yan B, Zhang HJ, Wang J, Xu G, Tang P, Duan W, Zhang SC. Large-gap quantum spin Hall insulators in tin films. Physical review letters. 2013 Sep 24;111(13):136804.

[56] Liu J, Qian X, Fu L. Crystal field effect induced topological crystalline insulators in monolayer IV–VI semiconductors. Nano letters. 2015 Mar 9;15(4):2657-61.

[57] Li L, Zhang X, Chen X, Zhao M. Giant topological nontrivial band gaps in chloridized gallium bismuthide. Nano letters. 2015 Jan 28;15(2):1296-301.

[58] Ugeda MM, Bradley AJ, Shi SF, Felipe H, Zhang Y, Qiu DY, Ruan W, Mo SK, Hussain Z, Shen ZX, Wang F. Giant bandgap renormalization and excitonic effects in a monolayer transition metal dichalcogenide semiconductor. Nature materials. 2014 Dec;13(12):1091.

[59] Klots AR, Newaz AK, Wang B, Prasai D, Krzyzanowska H, Lin J, Caudel D, Ghimire NJ, Yan J, Ivanov BL, Velizhanin KA. Probing excitonic states in suspended two-dimensional semiconductors by photocurrent spectroscopy. Scientific reports. 2014 Oct 16;4:6608.

[60] Zhu B, Chen X, Cui X. Exciton binding energy of monolayer $WS_2$. Scientific reports. 2015 Mar 18;5:9218.

[61] Hanbicki AT, Currie M, Kioseoglou G, Friedman AL, Jonker BT. Measurement of high exciton binding energy in the monolayer transition-metal dichalcogenides $WS_2$ and $WSe_2$. Solid State Communications. 2015 Feb 1;203:16-20.

[62] Khazaei M, Arai M, Sasaki T, Chung CY, Venkataramanan NS, Estili M, Sakka Y, Kawazoe Y. Novel electronic and magnetic properties of two-dimensional transition metal carbides and nitrides. Advanced Functional Materials. 2013 May 6;23(17):2185-92.

[63] Wang X, Jones AM, Seyler KL, Tran V, Jia Y, Zhao H, Wang H, Yang L, Xu X, Xia F. Highly anisotropic and robust excitons in monolayer black phosphorus. Nature nanotechnology. 2015 Jun;10(6):517.

[64] Zhang X. Excitonic Structure in Atomically-Thin Transition Metal Dichalcogenides (Doctoral dissertation, Columbia University).

[65] Plechinger G, Nagler P, Kraus J, Paradiso N, Strunk C, Schüller C, Korn T. Identification of excitons, trions and biexcitons in single-layer $WS_2$. physica status solidi (RRL)–Rapid Research Letters. 2015 Aug;9(8):457-61.

[66] Ceballos F, Bellus MZ, Chiu HY, Zhao H. Ultrafast charge separation and indirect exciton formation in a $MoS_2$–$MoSe_2$ van der Waals heterostructure. ACS nano. 2014 Nov 24;8(12):12717-24.

[67] Hong X, Kim J, Shi SF, Zhang Y, Jin C, Sun Y, Tongay S, Wu J, Zhang Y, Wang F. Ultrafast charge transfer in atomically thin $MoS_2/WS_2$ heterostructures. Nature nanotechnology. 2014 Sep;9(9):682.

[68] Kim J, Jin C, Chen B, Cai H, Zhao T, Lee P, Kahn S, Watanabe K, Taniguchi T, Tongay S, Crommie MF. Observation of ultralong valley lifetime in $WSe_2/MoS_2$ heterostructures. Science advances. 2017 Jul 1;3(7):e1700518.

[69] Jin C, Kim J, Utama MI, Regan EC, Kleemann H, Cai H, Shen Y, Shinner MJ, Sengupta A, Watanabe K, Taniguchi T. Imaging of pure spin-valley diffusion current in $WS_2$-$WSe_2$ heterostructures. Science. 2018 May 25;360(6391):893-6.

[70] Seyler KL, Rivera P, Yu H, Wilson NP, Ray EL, Mandrus DG, Yan J, Yao W, Xu X. Signatures of moiré-trapped valley excitons in $MoSe_2/WSe_2$ heterobilayers. Nature. 2019 Mar;567(7746):66.

[71] Tran K, Moody G, Wu F, Lu X, Choi J, Kim K, Rai A, Sanchez DA, Quan J, Singh A, Embley J. Evidence for moiré excitons in van der Waals heterostructures. Nature. 2019 Mar;567(7746):71.

[72] Jin C, Regan EC, Yan A, Utama MI, Wang D, Zhao S, Qin Y, Yang S, Zheng Z, Shi S, Watanabe K. Observation of moiré excitons in $WSe_2/WS_2$ heterostructure superlattices. Nature. 2019 Mar;567(7746):76.

[73] Alexeev EM, Ruiz-Tijerina DA, Danovich M, Hamer MJ, Terry DJ, Nayak PK, Ahn S, Pak S, Lee J, Sohn JI, Molas MR. Resonantly hybridized excitons in moiré superlattices in van der Waals heterostructures. Nature. 2019 Mar;567(7746):81.

[74] Zhang C, Chuu CP, Ren X, Li MY, Li LJ, Jin C, Chou MY, Shih CK. Interlayer couplings, Moiré patterns, and 2D electronic superlattices in $MoS_2/WSe_2$ hetero-bilayers. Science advances. 2017 Jan 1;3(1):e1601459.

[75] Rivera P, Schaibley JR, Jones AM, Ross JS, Wu S, Aivazian G, Klement P, Seyler K, Clark G, Ghimire NJ, Yan J. Observation of long-lived interlayer excitons in monolayer $MoSe_2$–$WSe_2$ heterostructures. Nature communications. 2015 Feb 24;6:6242.

[76] Merkl P, Mooshammer F, Steinleitner P, Girnghuber A, Lin KQ, Nagler P, Holler J, Schüller C, Lupton JM, Korn T, Ovesen S. Ultrafast transition between exciton phases in van der Waals heterostructures. Nature materials. 2019 Apr 8:1.

[77] Zhu H, Wang J, Gong Z, Kim YD, Hone J, Zhu XY. Interfacial charge transfer circumventing momentum mismatch at two-dimensional van der Waals heterojunctions. Nano letters. 2017 May 10;17(6):3591-8.

[78] Santos EJ, Kaxiras E. Electrically driven tuning of the dielectric constant in $MoS_2$ layers. ACS nano. 2013 Nov 25;7(12):10741-6.

[79] Mak KF, He K, Lee C, Lee GH, Hone J, Heinz TF, Shan J. Tightly bound trions in monolayer $MoS_2$. Nature materials. 2013 Mar;12(3):207.

[80] Ross JS, Wu S, Yu H, Ghimire NJ, Jones AM, Aivazian G, Yan J, Mandrus DG, Xiao D, Yao W, Xu X. Electrical control of neutral and charged excitons in a monolayer semiconductor. Nature communications. 2013 Feb 12;4:1474.

[81] Chernikov A, van der Zande AM, Hill HM, Rigosi AF, Velauthapillai A, Hone J, Heinz TF. Electrical tuning of exciton binding energies in monolayer $WS_2$. Physical review letters. 2015 Sep 16;115(12):126802.

[82] Wang K, De Greve K, Jauregui LA, Sushko A, High A, Zhou Y, Scuri G, Taniguchi T, Watanabe K, Lukin MD, Park H. Electrical control of charged carriers and excitons in atomically thin materials. Nature nanotechnology. 2018 Feb;13(2):128.

[83] Ciarrocchi A, Unuchek D, Avsar A, Watanabe K, Taniguchi T, Kis A. Polarization switching and electrical control of interlayer excitons in two-dimensional van der Waals heterostructures. Nature photonics. 2019 Feb;13(2):131.

[84] Aivazian G, Gong Z, Jones AM, Chu RL, Yan J, Mandrus DG, Zhang C, Cobden D, Yao W, Xu X. Magnetic control of valley pseudospin in monolayer $WSe_2$. Nature Physics. 2015 Feb;11(2):148.

[85] Li Y, Ludwig J, Low T, Chernikov A, Cui X, Arefe G, Kim YD, Van Der Zande AM, Rigosi A, Hill HM, Kim SH. Valley splitting and polarization by the Zeeman effect in monolayer MoSe 2. Physical review letters. 2014 Dec 23;113(26):266804.

[86] Srivastava A, Sidler M, Allain AV, Lembke DS, Kis A, Imamoğlu A. Valley Zeeman effect in elementary optical excitations of monolayer $WSe_2$. Nature Physics. 2015 Feb;11(2):141.

[87] MacNeill D, Heikes C, Mak KF, Anderson Z, Kormányos A, Zólyomi V, Park J, Ralph DC. Breaking of valley degeneracy by magnetic field in monolayer $MoSe_2$. Physical review letters. 2015 Jan 22;114(3):037401.

[88] Lyons TP, Dufferwiel S, Brooks M, Withers F, Taniguchi T, Watanabe K, Novoselov KS, Burkard G, Tartakovskii AI. The valley Zeeman effect in inter- and intra-valley trions in monolayer $WSe_2$. Nature communications. 2019 May 27;10(1):2330.

[89] Stier AV, Wilson NP, Clark G, Xu X, Crooker SA. Probing the influence of dielectric environment on excitons in monolayer $WSe_2$: insight from high magnetic fields. Nano letters. 2016 Oct 13;16(11):7054-60.

[90] Stevens CE, Paul J, Cox T, Sahoo PK, Gutiérrez HR, Turkowski V, Semenov D, McGill SA, Kapetanakis MD, Perakis IE, Hilton DJ. Biexcitons in monolayer transition metal dichalcogenides tuned by magnetic fields. Nature communications. 2018 Sep 13;9(1):3720.

[91] Sie EJ, McIver JW, Lee YH, Fu L, Kong J, Gedik N. Valley-selective optical Stark effect in monolayer $WS_2$. Nature materials. 2014 Dec 15;14: 290-294

[92] Kim J, Hong X, Jin C, Shi SF, Chang CY, Chiu MH, Li LJ, Wang F. Ultrafast generation of pseudo-magnetic field for valley excitons in $WSe_2$ monolayers. Science. 2014 Dec 5;346(6214):1205-8.

[93] Dufferwiel S, Lyons TP, Solnyshkov DD, Trichet AA, Withers F, Schwarz S, Malpuech G, Smith JM, Novoselov KS, Skolnick MS, Krizhanovskii DN. Valley-addressable polaritons in atomically thin semiconductors. Nature Photonics. 2017 Aug;11(8):497.

[94] Zeng H, Dai J, Yao W, Xiao D, Cui X. Valley polarization in $MoS_2$ monolayers by optical pumping. Nature nanotechnology. 2012 Aug;7(8):490.

[95] Sun Z, Gu J, Ghazaryan A, Shotan Z, Considine CR, Dollar M, Chakraborty B, Liu X, Ghaemi P, Kéna-Cohen S, Menon VM. Optical control of room-temperature valley polaritons. Nature Photonics. 2017 Aug;11(8):491.

[96] Chen YJ, Cain JD, Stanev TK, Dravid VP, Stern NP. Valley-polarized exciton–polaritons in a monolayer semiconductor. Nature Photonics. 2017 Jul;11(7):431.

[97] Kim SJ, Choi K, Lee B, Kim Y, Hong BH. Materials for flexible, stretchable electronics: graphene and 2D materials. Annual Review of Materials Research. 2015 Jul 1;45:63-84.

[98] Su J, He J, Zhang J, Lin Z, Chang J, Zhang J, Hao Y. Unusual properties and potential applications of strain $BN-MS_2$ (M=Mo, W) heterostructures. Scientific reports. 2019 Mar 5;9(1):3518.

[99] Defo RK, Fang S, Shirodkar SN, Tritsaris GA, Dimoulas A, Kaxiras E. Strain dependence of band gaps and exciton energies in pure and mixed transition-metal dichalcogenides. Physical Review B. 2016 Oct 27;94(15):155310.

[100] Kumar A, Ahluwalia PK. Mechanical strain dependent electronic and dielectric properties of two-dimensional honeycomb structures of MoX2 (X= S, Se, Te). Physica B: Condensed Matter. 2013 Jun 15;419:66-75.

[101] Hu Y, Zhang F, Titze M, Deng B, Li H, Cheng GJ. Straining effects in $MoS_2$ monolayer on nanostructured substrates: temperature-dependent photoluminescence and exciton dynamics. Nanoscale. 2018;10(12):5717-24.

[102] Arra S, Babar R, Kabir M. Exciton in phosphorene: Strain, impurity, thickness, and heterostructure. Physical Review B. 2019 Jan 22;99(4):045432.

[103] Aslan OB, Datye IM, Mleczko MJ, Sze Cheung K, Krylyuk S, Bruma A, Kalish I, Davydov AV, Pop E, Heinz TF. Probing the Optical Properties and Strain-Tuning of Ultrathin $Mo_{1−x}W_xTe_2$. Nano letters. 2018 Mar 21;18(4):2485-91.

[104] Feng J, Qian X, Huang CW, Li J. Strain-engineered artificial atom as a broad-spectrum solar energy funnel. Nature Photonics. 2012 Dec;6(12):866.

[105] Krustok J, Kaupmees R, Jaaniso R, Kiisk V, Sildos I, Li B, Gong Y. Local strain-induced band gap fluctuations and exciton localization in aged $WS_2$ monolayers. AIP Advances. 2017 Jun 5;7(6):065005.

[106] Castellanos-Gomez A, Roldán R, Cappelluti E, Buscema M, Guinea F, van der Zant HS, Steele GA. Local strain engineering in atomically thin $MoS_2$. Nano letters. 2013 Oct 3;13(11):5361-6.

[107] San-Jose P, Parente V, Guinea F, Roldán R, Prada E. Inverse funnel effect of excitons in strained black phosphorus. Physical Review X. 2016 Sep 27;6(3):031046.

[108] Wu S, Buckley S, Jones AM, Ross JS, Ghimire NJ, Yan J, Mandrus DG, Yao W, Hatami F, Vučković J, Majumdar A. Control of two-dimensional excitonic light emission via photonic crystal. 2D Materials. 2014 Apr 4;1(1):011001.

[109] Ye Y, Ye Z, Gharghi M, Zhu H, Zhao M, Wang Y, Yin X, Zhang X. Exciton-dominant electroluminescence from a diode of monolayer $MoS_2$. Applied Physics Letters. 2014 May 12;104(19):193508.

[110] Cheng R, Li D, Zhou H, Wang C, Yin A, Jiang S, Liu Y, Chen Y, Huang Y, Duan X. Electroluminescence and photocurrent generation from atomically sharp $WSe_2/MoS_2$ heterojunction p–n diodes. Nano letters. 2014 Sep 8;14(10):5590-7.

[111] Withers F, Del Pozo-Zamudio O, Mishchenko A, Rooney AP, Gholinia A, Watanabe K, Taniguchi T, Haigh SJ, Geim AK, Tartakovskii AI, Novoselov KS. Light-emitting diodes by band-structure engineering in van der Waals heterostructures. Nature materials. 2015 Mar;14(3):301.

[112] Wang S, Wang J, Zhao W, Giustiniano F, Chu L, Verzhbitskiy I, Zhou Yong J, Eda G. Efficient carrier-to-exciton conversion in field emission tunnel diodes based on MIS-type van der Waals heterostack. Nano letters. 2017 Jul 24;17(8):5156-62.

[113] Sundaram RS, Engel M, Lombardo A, Krupke R, Ferrari AC, Avouris P, Steiner M. Electroluminescence in single layer $MoS_2$. Nano letters. 2013 Mar 29;13(4):1416-21.

[114] Kim YD, Gao Y, Shiue RJ, Wang L, Aslan OB, Bae MH, Kim H, Seo D, Choi HJ, Kim SH, Nemilentsau A. Ultrafast graphene light emitters. Nano letters. 2018 Jan 22;18(2):934-40.

[115] Ross JS, Rivera P, Schaibley J, Lee-Wong E, Yu H, Taniguchi T, Watanabe K, Yan J, Mandrus D, Cobden D, Yao W. Interlayer exciton optoelectronics in a 2D heterostructure p–n junction. Nano letters. 2017 Jan 4;17(2):638-43.

[116] Lee CH, Lee GH, Van Der Zande AM, Chen W, Li Y, Han M, Cui X, Arefe G, Nuckolls C, Heinz TF, Guo J. Atomically thin p–n junctions with van der Waals heterointerfaces. Nature nanotechnology. 2014 Sep;9(9):676.

[117] Furchi MM, Pospischil A, Libisch F, Burgdörfer J, Mueller T. Photovoltaic effect in an electrically tunable van der Waals heterojunction. Nano letters. 2014 Jul 28;14(8):4785-91.

[118] Qiu H, Xu T, Wang Z, Ren W, Nan H, Ni Z, Chen Q, Yuan S, Miao F, Song F, Long G. Hopping transport through defect-induced localized states in molybdenum disulphide. Nature communications. 2013 Oct 23;4:2642.

[119] Schmidt H, Giustiniano F, Eda G. Electronic transport properties of transition metal dichalcogenide field-effect devices: surface and interface effects. Chemical Society Reviews. 2015;44(21):7715-36.

[120] Withers F, Del Pozo-Zamudio O, Schwarz S, Dufferwiel S, Walker PM, Godde T, Rooney AP, Gholinia A, Woods CR, Blake P, Haigh SJ. $WSe_2$ light-emitting tunneling transistors with enhanced brightness at room temperature. Nano letters. 2015 Nov 16;15(12):8223-8.

[121] Li D, Cheng R, Zhou H, Wang C, Yin A, Chen Y, Weiss NO, Huang Y, Duan X. Electric-field-induced strong enhancement of electroluminescence in multilayer molybdenum disulfide. Nature communications. 2015 Jul 1;6:7509.

[122] Jariwala D, Davoyan AR, Wong J, Atwater HA. Van der Waals materials for atomically-thin photovoltaics: promise and outlook. Acs Photonics. 2017 Nov 1;4(12):2962-70.

[123] Ganesan VD, Linghu J, Zhang C, Feng YP, Shen L. Heterostructures of phosphorene and transition metal dichalcogenides for excitonic solar cells: A first-principles study. Applied Physics Letters. 2016 Mar 21;108(12):122105.

[124] Jang MS, Brar VW, Sherrott MC, Lopez JJ, Kim LK, Kim S, Choi M, Atwater HA. Tunable Large Resonant Absorption in a Mid-IR Graphene Salisbury Screen. arXiv preprint arXiv:1312.6463. 2013 Dec 23.

[125] Aydin K, Ferry VE, Briggs RM, Atwater HA. Broadband polarization-independent resonant light absorption using ultrathin plasmonic super absorbers. Nature communications. 2011 Nov 1;2:517.

[126] Piper JR, Fan S. Total absorption in a graphene monolayer in the optical regime by critical coupling with a photonic crystal guided resonance. Acs Photonics. 2014 Mar 15;1(4):347-53.

[127] Kim SJ, Fan P, Kang JH, Brongersma ML. Creating semiconductor metafilms with designer absorption spectra. Nature communications. 2015 Jul 17;6:7591.

[128] Mak KF, Lee C, Hone J, Shan J, Heinz TF. Atomically thin $MoS_2$: a new direct-gap semiconductor. Physical review letters. 2010 Sep 24;105(13):136805.

[129] Jariwala D, Davoyan AR, Tagliabue G, Sherrott MC, Wong J, Atwater HA. Near-unity absorption in van der Waals semiconductors for ultrathin optoelectronics. Nano letters. 2016 Aug 31;16(9):5482-7.

[130] Wong J, Jariwala D, Tagliabue G, Tat K, Davoyan AR, Sherrott MC, Atwater HA. High photovoltaic quantum efficiency in ultrathin van der Waals heterostructures. ACS nano. 2017 Jun 7;11(7):7230-40.

[131] Bahauddin SM, Robatjazi H, Thomann I. Broadband absorption engineering to enhance light absorption in monolayer $MoS_2$. Acs Photonics. 2016 May 4;3(5):853-62.

[132] Sze SM. Semiconductor devices: physics and technology. John wiley & sons; 2008.

[133] Konstantatos G, Sargent EH. Nanostructured materials for photon detection. Nature nanotechnology. 2010 Jun;5(6):391.

[134] Koppens FH, Mueller T, Avouris P, Ferrari AC, Vitiello MS, Polini M. Photodetectors based on graphene, other two-dimensional materials and hybrid systems. Nature nanotechnology. 2014 Oct;9(10):780.

[135] Sun Z, Chang H. Graphene and graphene-like two-dimensional materials in photodetection: mechanisms and methodology. ACS nano. 2014 Apr 14;8(5):4133-56.

[136] Xia F, Mueller T, Lin YM, Valdes-Garcia A, Avouris P. Ultrafast graphene photodetector. Nature nanotechnology. 2009 Dec;4(12):839.

[137] Konstantatos G. Current status and technological prospect of photodetectors based on two-dimensional materials. Nature communications. 2018 Dec 10;9(1):5266.

[138] Lopez-Sanchez O, Lembke D, Kayci M, Radenovic A, Kis A. Ultrasensitive photodetectors based on monolayer $MoS_2$. Nature nanotechnology. 2013 Jul;8(7):497.

[139] Konstantatos G, Badioli M, Gaudreau L, Osmond J, Bernechea M, De Arquer FP, Gatti F, Koppens FH. Hybrid graphene–quantum dot phototransistors with ultrahigh gain. Nature nanotechnology. 2012 Jun;7(6):363.

[140] Lee Y, Kwon J, Hwang E, Ra CH, Yoo WJ, Ahn JH, Park JH, Cho JH. High-performance perovskite–graphene hybrid photodetector. Advanced materials. 2015 Jan;27(1):41-6.

[141] Chen Z, Cheng Z, Wang J, Wan X, Shu C, Tsang HK, Ho HP, Xu JB. High responsivity, broadband, and fast graphene/silicon photodetector in photoconductor mode. Advanced Optical Materials. 2015 Sep;3(9):1207-14.

[142] Liu Y, Wang F, Wang X, Wang X, Flahaut E, Liu X, Li Y, Wang X, Xu Y, Shi Y, Zhang R. Planar carbon nanotube–graphene hybrid films for high-performance broadband photodetectors. Nature communications. 2015 Oct 8;6:8589.

[143] Tao L, Chen Z, Li X, Yan K, Xu JB. Hybrid graphene tunneling photoconductor with interface engineering towards fast photoresponse and high responsivity. npj 2D Materials and Applications. 2017 Jul 3;1(1):19.

[144] Park J, Ahn YH, Ruiz-Vargas C. Imaging of photocurrent generation and collection in single-layer graphene. Nano letters. 2009 Mar 27;9(5):1742-6.

[145] Xu X, Gabor NM, Alden JS, van der Zande AM, McEuen PL. Photo-thermoelectric effect at a graphene interface junction. Nano letters. 2009 Nov 10;10(2):562-6.

[146] Miao X, Tongay S, Petterson MK, Berke K, Rinzler AG, Appleton BR, Hebard AF. High efficiency graphene solar cells by chemical doping. Nano letters. 2012 May 10;12(6):2745-50.

[147] An X, Liu F, Jung YJ, Kar S. Tunable graphene–silicon heterojunctions for ultrasensitive photodetection. Nano letters. 2013 Feb 5;13(3):909-16.

[148] Lopez-Sanchez O, Lembke D, Kayci M, Radenovic A, Kis A. Ultrasensitive photodetectors based on monolayer $MoS_2$. Nature nanotechnology. 2013 Jul;8(7):497.

[149] Chernikov A, Berkelbach TC, Hill HM, Rigosi A, Li Y, Aslan OB, Reichman DR, Hybertsen MS, Heinz TF. Exciton binding energy and nonhydrogenic Rydberg series in monolayer $WS_2$. Physical review letters. 2014 Aug 13;113(7):076802.

[150] Mak KF, McGill KL, Park J, McEuen PL. The valley Hall effect in $MoS_2$ transistors. Science. 2014 Jun 27;344(6191):1489-92.

[151] Tsai DS, Liu KK, Lien DH, Tsai ML, Kang CF, Lin CA, Li LJ, He JH. Few-layer $MoS_2$ with high broadband photogain and fast optical switching for use in harsh environments. Acs Nano. 2013 Apr 22;7(5):3905-11.

[152] Yu WJ, Liu Y, Zhou H, Yin A, Li Z, Huang Y, Duan X. Highly efficient gate-tunable photocurrent generation in vertical heterostructures of layered materials. Nature nanotechnology. 2013 Dec;8(12):952.

[153] Furchi MM, Polyushkin DK, Pospischil A, Mueller T. Mechanisms of photoconductivity in atomically thin $MoS_2$. Nano letters. 2014 Oct 13;14(11):6165-70.

[154] Lee CH, Lee GH, Van Der Zande AM, Chen W, Li Y, Han M, Cui X, Arefe G, Nuckolls C, Heinz TF, Guo J. Atomically thin p–n junctions with van der Waals heterointerfaces. Nature nanotechnology. 2014 Sep;9(9):676.

[155] Massicotte M, Schmidt P, Vialla F, Schädler KG, Reserbat-Plantey A, Watanabe K, Taniguchi T, Tielrooij KJ, Koppens FH. Picosecond photoresponse in van der Waals heterostructures. Nature nanotechnology. 2016 Jan;11(1):42.

[156] Ross JS, Klement P, Jones AM, Ghimire NJ, Yan J, Mandrus DG, Taniguchi T, Watanabe K, Kitamura K, Yao W, Cobden DH. Electrically tunable excitonic light-emitting diodes based on monolayer $WSe_2$ p–n junctions. Nature nanotechnology. 2014 Apr;9(4):268.

## Figure Captions

**Figure 1.** The gallery of 2D materials

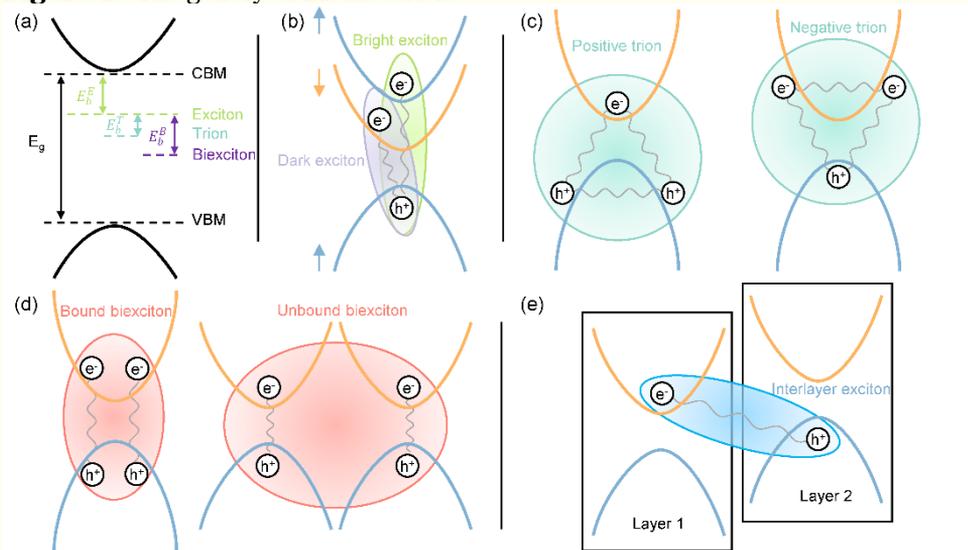

**Figure 2.** Different exciton types in atomically thin nanomaterials and related heterostructures. (**a**) The schematic for the energy level. (**b**) Excitons are Coulomb-bound electron hole pairs (ovals in the picture): bright excitons consist of electrons and holes with antiparallel spins, while dark excitons consist of electrons and holes with parallel spins. (**c**) Trions emerge when an additional electron (hole) joins the exciton. (**d**) Biexcitons are created from two free excitons with different total momenta. (**e**) Interlayer excitons appear when electrons and holes are located in different layers.

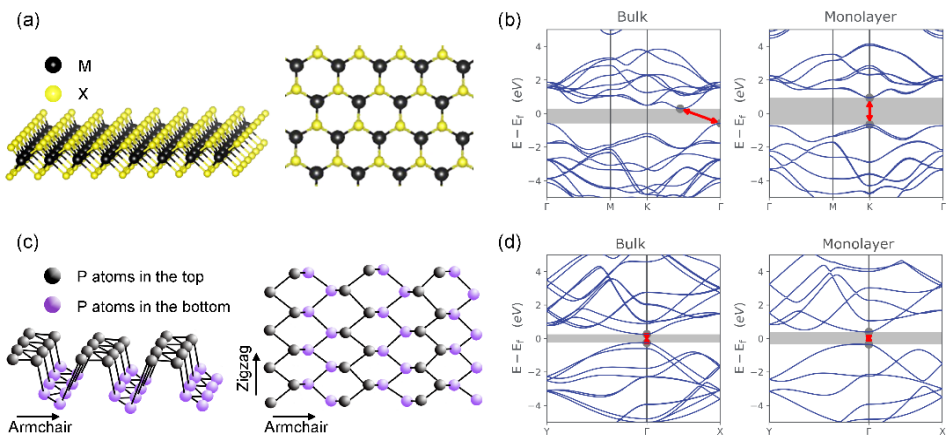

**Figure 3.** Atomic structures and electronic structures of TMDs and phosphorene: side view (left) and top view (right) of the atomic structures of the monolayer semiconducting 2H-phase TMDs (**a**) and of the monolayer phosphorene (**c**); band structures of bulk and monolayer $MoS_2$ (**b**) and phosphorene (**d**) [45]. Note that the bandgap shows a widening in phosphorene and both a widening and a crossover from indirect to direct bandgap in $MoS_2$. Reproduced with permission [45]. Copyright 2019 Ossila Ltd.

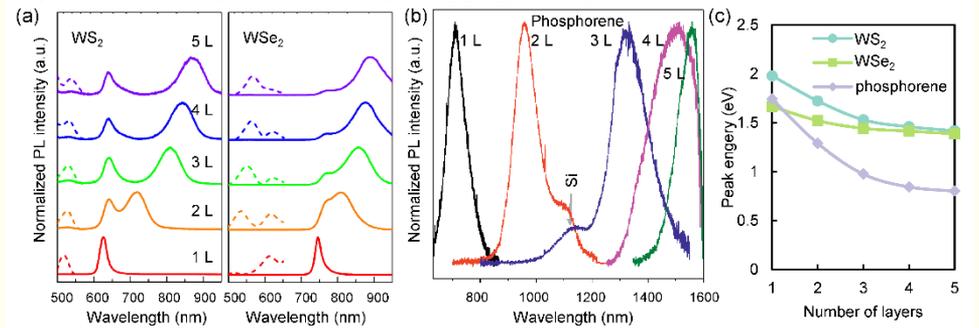

**Figure 4.** The effects of layer number on the PL spectra and peak energy of TMDs and phosphorene. (**a, b**) Normalized PL spectra of $2H$-$WS_2$, $2H$-$WSe_2$ and phosphorene flakes consisting of 1-5 layers. Each PL spectra is normalized to its peak intensity and system background [31, 47]. (**c**) Evolution of PL peak energy with layer number of $2H$-$WS_2$, $2H$-$WSe_2$, and phosphorene from (**a, b**), showing an increase in peak energy as the layer number reduces. (**a**) Reproduced with permission [47]. Copyright 2012 American Chemical Society. (**b**) Reproduced with permission [31]. Copyright 2015 Springer Nature Publishing AG.

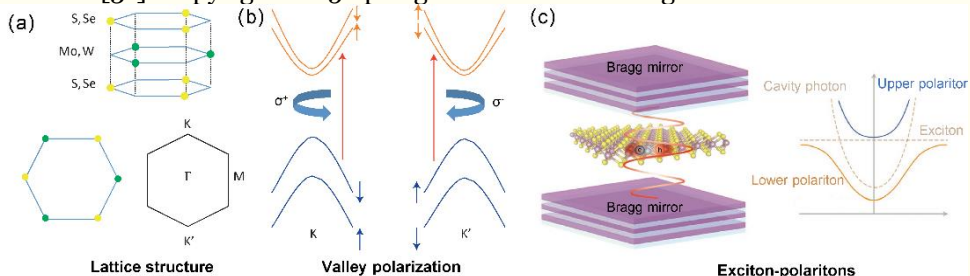

**Figure 5.** Lattice structure, valley polarization, and exciton-polaritons in 2D TMDs. (**a**) The honeycomb lattice structure of monolayer TMDs, with broken inversion symmetry and the high-symmetry points in the first Brillouin zone. (**b**) Electronic bands around the K and K' points, which are spin-split by the spin–orbit interactions. The spin (up and down arrows) and valley (K and K') degrees of freedom are locked together. (**c**) Exciton–polariton states in a 2D semiconductor

embedded inside a photonic microcavity. (**a, b**) Reproduced with permission [49]. Copyright 2016 Springer Nature Publishing AG. (**c**) Reproduced with permission [50]. Copyright 2019 John Wiley & Sons, Inc.

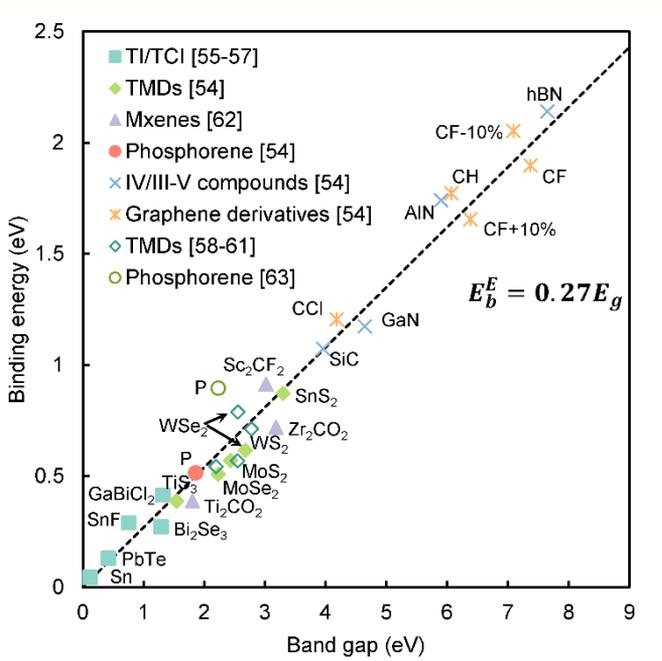

**Figure 6.** Linear relationship between quasiparticle bandgap ($E_g$) and exciton binding energy ($E_b^E$). Reproduced with permission [54]. Copyright 2017 American Physical Society.

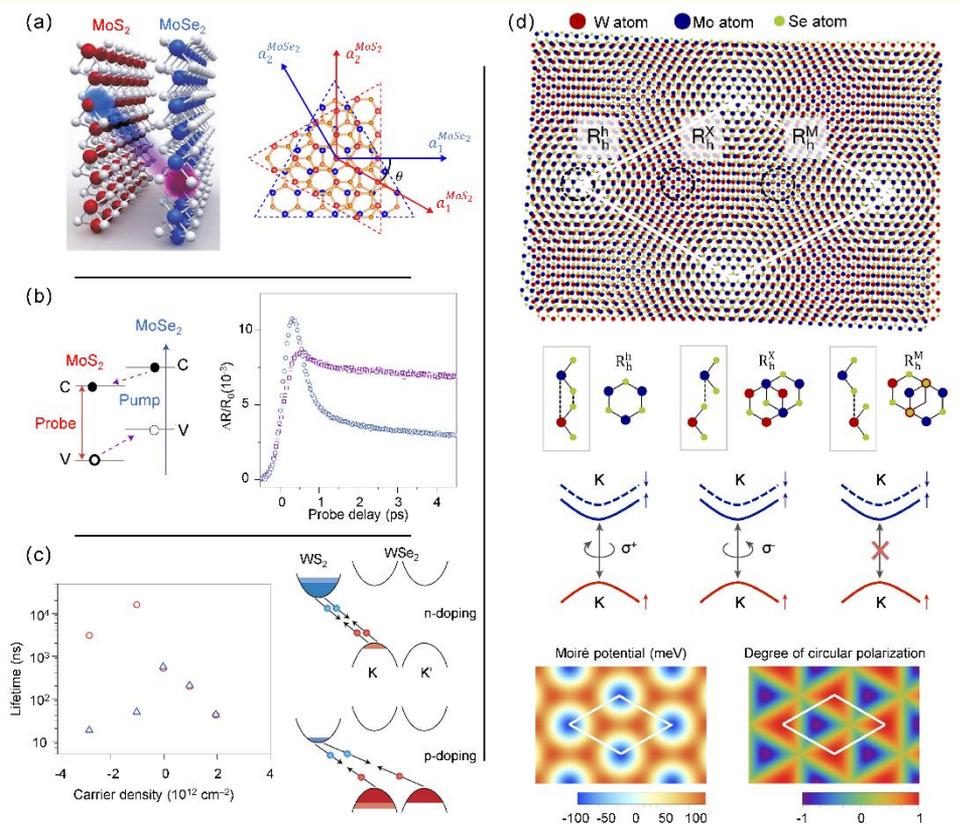

**Figure 7.** Excitonic effects in vdWHs. (**a**) Sketch of $MoS_2/MoSe_2$ heterobilayer (left) and its moiré superlattice (right) [8]. (**b**) Schematic of a pump-probe configuration (left), and time-resolved differential reflection of a $MoS_2/MoSe_2$ heterobilayer (blue) and of $MoS_2$ monolayer (purple) (right) [66]. (**c**) Comparison between spin-valley lifetime (circles) and hole population lifetime (triangles) under different carrier concentration in $MoS_2/MoSe_2$ heterostructure (left), and Schematic illustration of the interlayer electron–hole recombination process in electron-doped and hole-doped heterostructures [69]. (**d**) Moiré superlattice modulates the electronic and optical properties in $WSe_2/MoSe_2$ heterostructure: three different local atomic alignments and their corresponding schematic (top), the moiré potential of the interlayer exciton transition (left lower), and spatial map of the optical selection rules for K-valley excitons (right lower) [71]. (**a**) Reproduced with permission [8]. Copyright 2016 American Association for the Advancement of Science. (**b**) Reproduced with permission [66]. Copyright 2014, American Chemical Society. (**c**) Reproduced with permission [69]. Copyright 2018 American Association for the Advancement of Science. (**d**) Reproduced with permission [71]. Copyright 2019 Springer Nature Publishing AG.

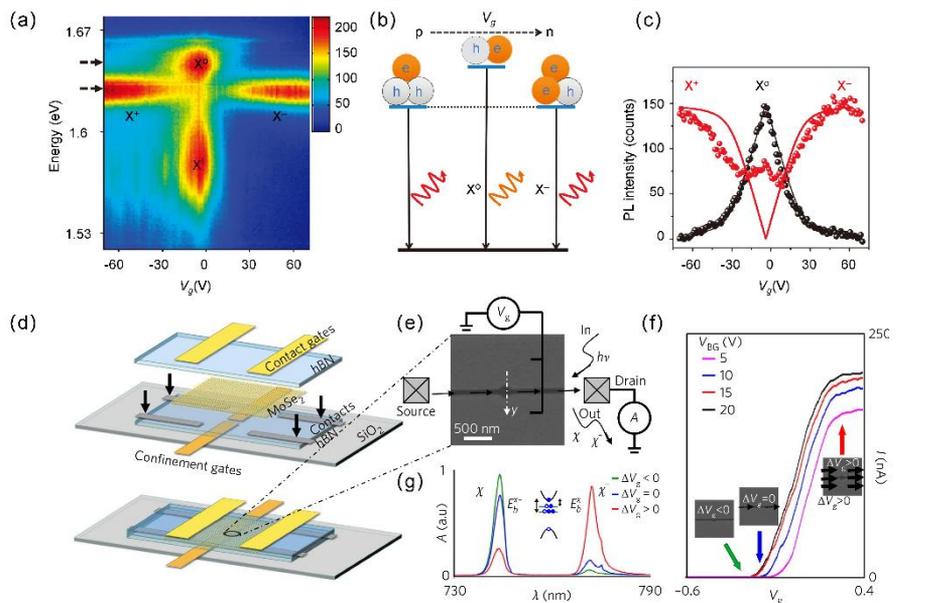

**Figure 8.** Electrical tuning of excitons. (**a-c**) Electrical control in monolayer 2D materials [80]: (**a**) MoSe$_2$ PL is plotted as a function of back-gate voltage, showing a transition from positive trion to negative trion as gate voltage increases. (b) Illustration of the gate-dependent transitions and quasiparticles. (**c**) The relationship between trion and exciton peak intensity and gate voltage at dashed arrows in (**a**). Solid lines are fits based on the mass action model. (**d-g**) Electrical control in vdWHs [82]: (d) Optoelectronic transport device consisting of hBN/MoSe$_2$/hBN heterostructure. (**e**) SEM image of a gate-defined monolayer MoSe$_2$ quantum dot. (**f**) Typically measured current across the device as a function of local gate voltage $V_g$ at different silicon backgate voltage VBG. (**g**) recombination emission signals of excitons and trions as a function of emission wavelength at different $V_g$ values. (**a-c**) Reproduced with permission [80]. Copyright 2013 Springer Nature Publishing AG. (**d-g**) Reproduced with permission [82]. Copyright 2018 Springer Nature Publishing AG.

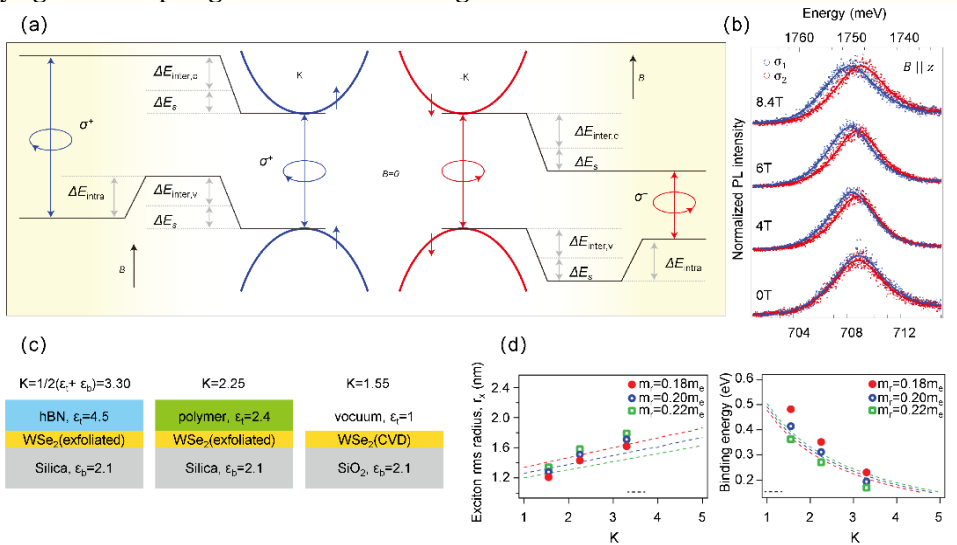

**Figure 9.** Magnetic tuning of excitons. (**a, b**) Valley Zeeman effect [86]. (**a**)Valley Zeeman effect in a finite out-of-plane B, the degeneracy between the ±K valleys is attributed to three factors: the spin-Zeeman effect ($ΔE_s$), the intercellular

orbital magnetic moment ($\Delta E_{inter}$), and the intracellular contribution from the $d \pm id$ orbitals of the valence band ($\Delta E_{intra}$). The signs of these contributions are opposite in the two valleys. (**b**) Normalized polarization-resolved PL spectra of the neutral exciton peak as a function of the out-of-plane magnetic field (B), indicating a B-dependent splitting phenomenon via valley Zeeman effect. (**c, d**) Electrical control by surrounding dielectric environment [89]: (**c**) The surrounding dielectric environments are changing by encapsulating hBN, polymer, or nothing on WSe$_2$ monolayer on silica substrate, where the average dielectric constant is defined as $k=(\varepsilon_t+\varepsilon_b)/2$ ($\varepsilon_t$ and $\varepsilon_b$ are the relative dielectric constants of the bottom substrate and the top encapsulation overlayer, respectively). (**d**) exciton root-mean-square (rms) radius $r_X$ and exciton binding energy as a function of $k$ (points and lines are the results from experiments and screened Keldysh model, respectively), where $m_e$, $m_r$, and $r_o$ are the exciton mass, the reduced mass of the exciton, and the characteristic screening length, respectively. (**a, b**) Reproduced with permission [86]. Copyright 2013 Springer Nature Publishing AG. (**c, d**) Reproduced with permission [89]. Copyright 2016 American Physical Society.

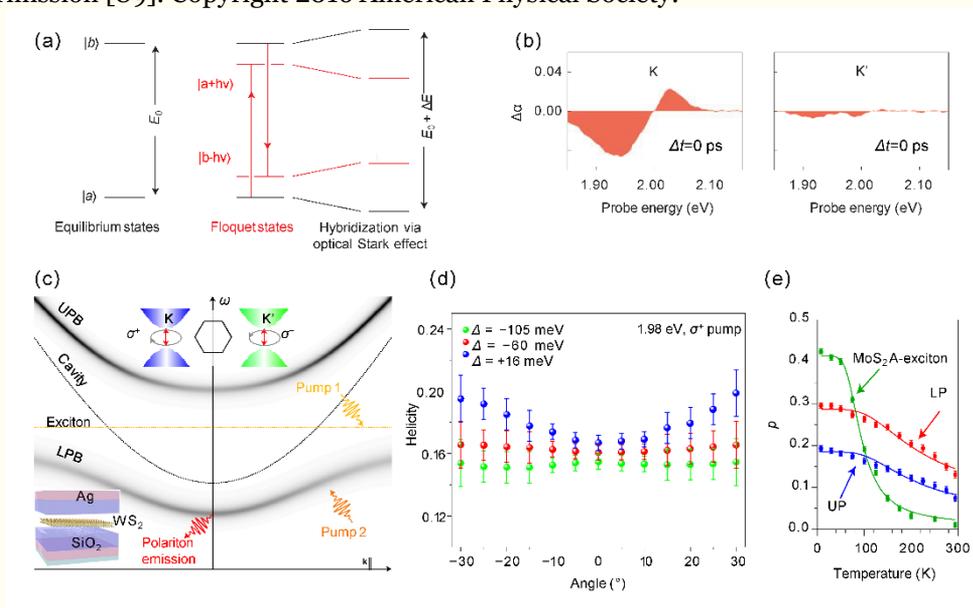

**Figure 10.** Optical tuning of excitons. (**a, b**) optical Stark effect [91]. (**a**) Illustration of optical stark effect for two-level system. Ground state |a⟩ and excited state |b⟩ can hybridize with Floquet states |a + ℏω⟩ and |b + ℏω⟩, bringing in shifted energy levels. (**b**) The valley selectivity of the optical Stark effect, showing an effect only at K valley by σ – polarization pump pulses. (**c~e**) Valley polaritons via optical pumping [95,96]. (**c**) Schematic of the valley polariton phenomena. The lower polariton branch (LBP) and the upper polariton branch (UPB) are the solid curves. The valley-polarization phenomena, caused by the broken inversion symmetry, is inserted in the top. (**d**) Polariton emission with angle-dependent helicity. Angle-resolved helicity was measured for three detuned cavities Δ at the σ+ excitation, where only the positive detuned cavities shows increasing helicity as a function of angle. (**e**) Exciton-polaritons with a temperature-dependent emission polarization. Emission polarization for bare exciton, and upper polariton (UP) and lower polariton (LP) branches change with temperature. (**a, b**) Reproduced with permission [91]. Copyright 2014 Springer Nature Publishing AG. (**c, d**) Reproduced with permission [95]. Copyright 2017 Springer Nature Publishing AG. (**e**) Reproduced with permission [96]. Copyright 2017 Springer Nature Publishing AG.

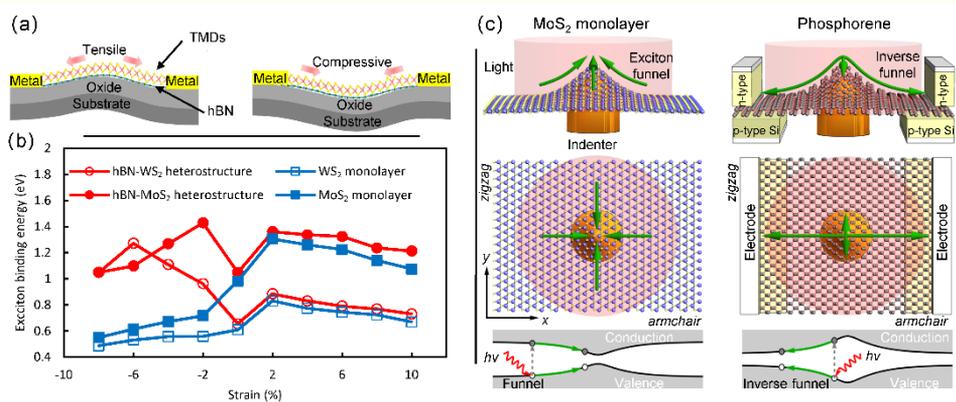

**Figure 11.** Mechanical tuning of excitons. (**a, b**) Exciton binding energies under compressive and tensile strain [98]. (**a**) Schematics of hBN-TMDs heterostructures nanodevices with tensile and compressive strain. (**b**) Exciton binding energies of TMD monolayer and hBN-TMD heterostructures as functions of strain. (**c**) Funnel Effect of Excitons under indentation. When an indenter creates an inhomogeneous strain profile that modulates the gap, excitons (in green) in $MoS_2$ concentrate on isotropically the center, while excitons in phosphorene disperse, especially along the armchair direction [107]. (**a, b**) Reproduced with permission [98]. Copyright 2019 Springer Nature Publishing AG. (**c**) Reproduced with permission [107]. Copyright 2016 American Physical Society.

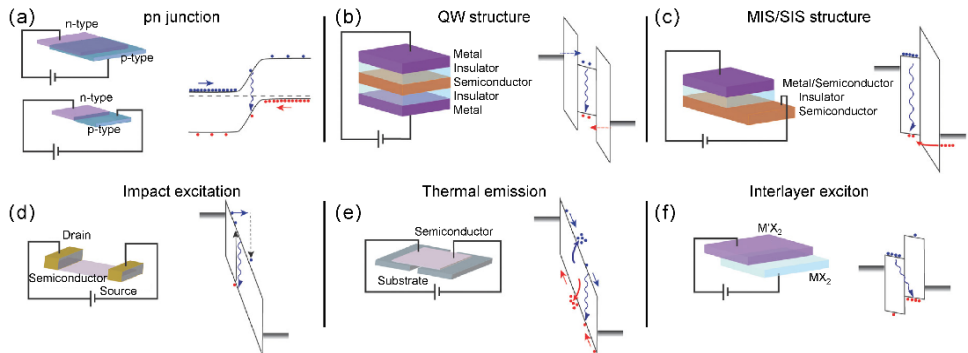

**Figure 12.** EL device structures and emission mechanisms [49]. (**a**) Vertical and lateral p-n junctions. (**b**) Quantum well heterojunction structure. (**c**) Metal–insulator–semiconductor (MIS) and semiconductor–insulator–semiconductor (SIS) structure. (**d**) Lateral unipolar device where emission is induced by impact excitation. (**e**) Locally suspended thermal emission device. (**f**) Hetero-bilayer device exhibiting interlayer exciton emission. Reproduced with permission [49]. Copyright 2016 Springer Nature Publishing AG.

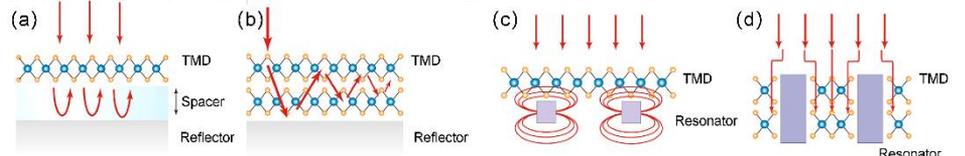

**Figure 13.** Possible light trapping configurations for enhancing sunlight absorption for Photovoltaics [122]. (**a**) Salisbury screen-like configuration where a spacer with ~λ/4 thickness sandwichs between a low loss metal reflector and a monolayer absorber. (**b**) Multilayer vdWH absorber directly placed on a smooth reflective metal reflector. (**c**) TMD monolayer coupled with resonators/antennas. (**d**) Multilayer vdWH absorber etched by nanometer scale antennas/resonators. Reproduced with permission [122]. Copyright 2017 American Chemical Society.

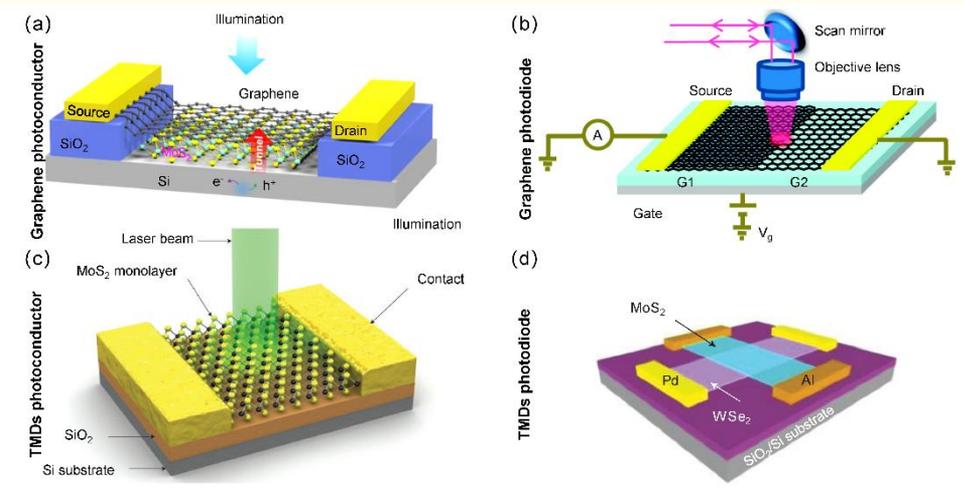

**Figure 14.** Typical 2D photodetectors. (**a**) Schematic of a hybrid graphene photoconductor [143]. (**b**) Schematic of a single-bilayer graphene interface junction, in which photocurrent generation is dominant by photothermoelectric effect [145]. (**c**) Schematic of monolayer $MoS_2$ lateral photoconductor [148]. (**d**) Schematic of vertical p–n photodiode formed by monolayer $MoS_2$ and $WSe_2$, in which a photocurrent hot spot is produced at the heterojunction [154]. (**a**) Reproduced with permission [143]. Copyright 2017 Springer Nature Publishing AG. (**b**) Reproduced with permission [145]. Copyright 2009 American Chemical Society. (**c**) Reproduced with permission [148]. Copyright 2013 Springer Nature Publishing AG. (**d**) Reproduced with permission [154]. Copyright 2014 Springer Nature Publishing AG.